\documentclass[aps,english,showpacs,twocolumn,prl]{revtex4-1}
\usepackage{amsfonts}
\usepackage{amssymb}
\usepackage{amsmath}
\usepackage{graphicx}
\usepackage{epsfig}

\begin{document}

\title{Non-Hermitian Aharonov-Bohm Cage}
\author{S. M. Zhang}
\author{L. Jin}
\email{jinliang@nankai.edu.cn}
\affiliation{School of Physics, Nankai University, Tianjin 300071, China}

\begin{abstract}
Aharonov-Bohm (AB) cage has a spectrum fully constituted by the flat bands
and has the capacity to confine arbitrary excitation. The confined
excitation in coupled waveguides propagates without diffraction. Exploited
the unidirectionality of the exceptional point (EP), we first propose the
non-Hermitian AB cage in the photonic crystal with inversion symmetry or
combined-inversion symmetry. Alternatively, the destructive interference of
the synthetic magnetic flux incorporated with gain and loss at the EP is
then shown to be able to induce the non-Hermitian AB cage. The spectrum of
the non-Hermitian AB cage is entirely constituted by the coalesced flat
bands. The light excitation is still confined regardless of the nonunitary
dynamics caused by the non-Hermiticity, but the localization area may alter.
The non-Hermitian AB caging can also be observed in the passive photonic
crystals including coupled waveguides. These findings pave the way of light
control and manipulation in the non-Hermitian integrated photonics.
\end{abstract}

\maketitle

\textit{Introduction.---}Light flow engineering is fundamentally important
in optics. The confinement of light is an important task for light control
and manipulation. The flat bands are dispersiveless, support the compact
localized state (CLS), and play a pivotal role for the light confinement. A
vast of one-dimensional (1D) and two-dimensional (2D) lattices support flat
bands \cite%
{CEC,Yamamoto,Leykam13,SFlachPRL113,LMI,Jacqmin,FlachBO,Baboux,FlachPRB17,SDHuber,LeykamAPX}
and have been demonstrated experimentally in the direct femtosecond laser
writing of optical waveguides \cite%
{Mukherjee15,Mukherjee17,LiebNJP,MolinaPRL,ThomsonPRL2015}. In addition, the
photonic system serves as an ideal platform for the study of non-Hermitian
physics \cite%
{PTRevLGe,PTRev,Longhi17,Makris,Gupta,Christodoulides,Ozdemir,Miri}. Flat
bands in non-Hermitian photonic crystals are successfully proposed \cite%
{Molina,RamezaniFB,LeykamFB,LGePR,LGeLieb,LGePRL2018,SzameitPRL19,JLPRA}.
The $\mathcal{PT}$-symmetric real-energy flat band is observed at the band
gap closure of a non-Hermitian triangular coupled waveguide lattice \cite%
{RamezaniFB,SzameitPRL19}.

In contrast to the general confinement of flat-band CLS, an Aharonov-Bohm
(AB) cage completely confines the light excitation that is not limited to
the CLS of the system. This is a consequence that the spectrum of AB cage is
fully consisted of flat bands \cite{Vidal,Vidal01}. Recently, a rhombic
lattice with a half quantum magnetic flux in each plaquette forms an AB cage
and has been realized experimentally in the coupled waveguides \cite%
{SLOL,Goldman18,ASarXiv,Maluckov,DiLiberto}. Here, we exploit the
exceptional point (EP) to propose non-Hermitian AB cages in the 1D and 2D
photonic resonator lattices, where all the bands are flat and consisting of
coalesced energy levels. The oppositely orientated unidirectional couplings
or the destructive interference at the proper match between synthetic
magnetic flux and balanced gain and loss at the EP induces the non-Hermitian
AB caging. Although the band energies are entirely real and arbitrary
excitation is completely confined, the excitation intensities in the
non-Hermitian AB cages can be constant, oscillate or polynomially increase
with time and the localization area of excitation may alter.

\textit{Confinement by unidirectional coupling.---}The EP provides many
potential applications \cite{Miri} including optical sensing \cite%
{Wiersig,NMNJP,ZPLiu,EP2Sensing,EP3Sensing,HKLau,WChen18,MZhang,CC,Djorwe,YHLai,MPH}%
, power transfer \cite{HXu,SFanNature}, and unidirectional lasing \cite%
{Ramezani14,LYangPNAS,LJinPRL}. The perfect unidirectionality is an
intrinsic feature of the EP. The nonreciprocal (asymmetric) coupling
strength results in one-way amplification or attenuation; an extreme case is
the unidirectional coupling at the EP, where tunneling is one-way allowed
and photons in one site can tunnel to the other site with the prohibition of
the reverse process. This inspires the design of non-Hermitian AB cages as a
novel application of the EP for the complete light confinement by
alternatively introducing the unidirectional couplings. Arbitrary light
excitation will be confined inside the area with only inward unidirectional
couplings connected to the outside.

In Hermitian optical structures, the nonreciprocity can be generated through
dynamical modulation \cite{Alu17}, optomechanical coupling \cite{Miri17}, or
optical nonlinearity in asymmetric systems \cite{Qi12}. In non-Hermitian
optical structures, the gain and loss associated with synthetic magnetic
flux create the nonreciprocity \cite{LJinPRL}. The external pumping induces
gain \cite{Jean,Parto,Feng}; intentionally introduced absorption induces
loss. For example, fabricating the radiative loss \cite{Poli,Weimann,MPan}
or cutting the waveguide \cite{Rechtsman19} generates extra loss in coupled
waveguides; while sticking additional material with strong absorption
generates extra loss in coupled resonators \cite{Poli}. Besides, the
asymmetric coupling has been proposed in coupled resonator array by imposing
an effective imaginary magnetic field \cite{LonghiPRBSR}. Both approaches
enable the nonreciprocity and optical diode in non-Hermitian metamaterial
\cite{Jalas}.

\begin{figure}[tb]
\includegraphics[ bb=0 0 605 495, width=8.8 cm, clip]{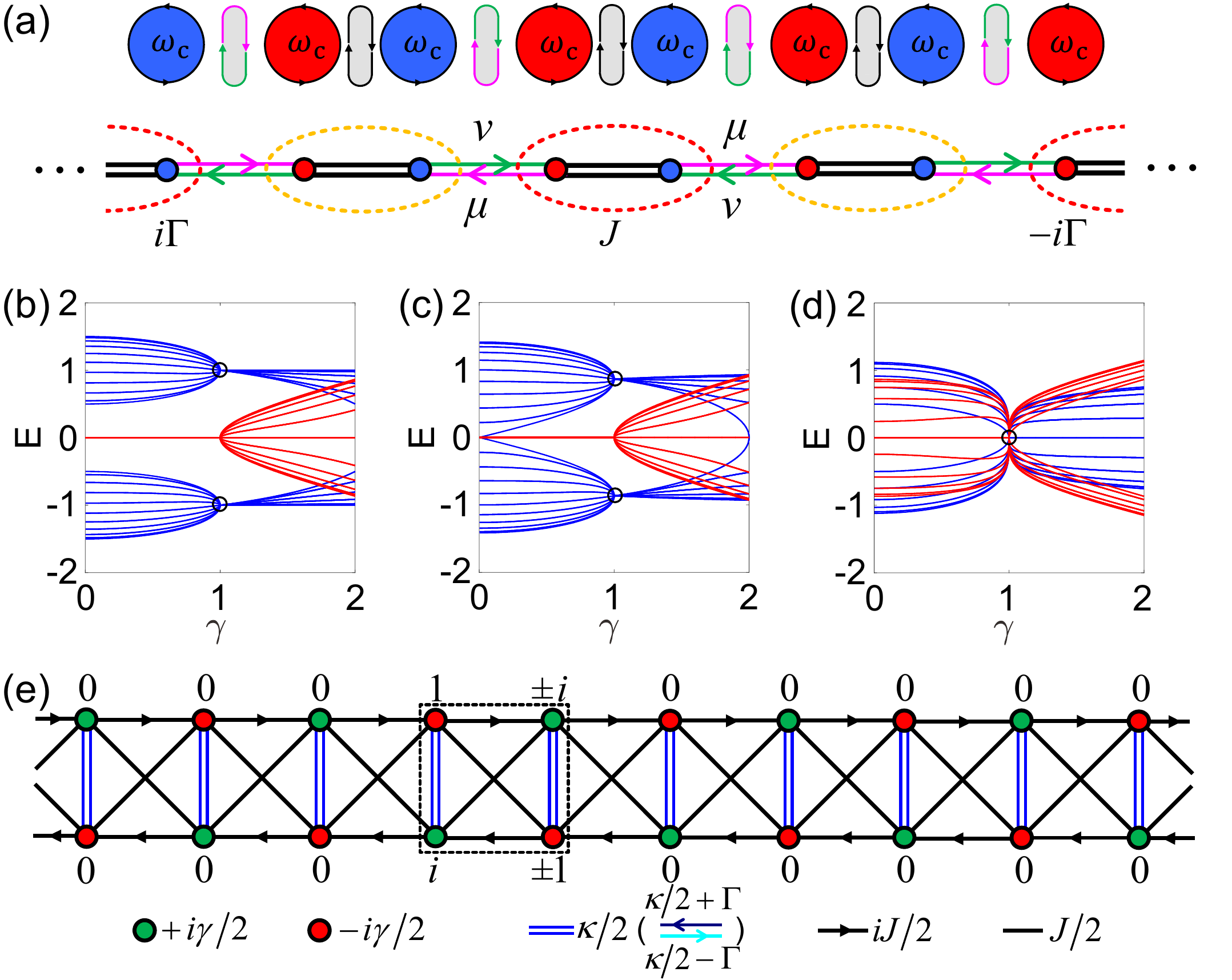}
\caption{(a) Schematic of the 1D non-Hermitian coupled resonator array,
which is inversion symmetric (chiral-inversion symmetric) when $\Gamma=0$ ($\Gamma \neq 0$). Spectra of (a) at (b) $\Gamma=0$, (c) $\Gamma=1/2$, and (d)
$\Gamma=1$ with the real (imaginary) part in blue (red); the zero energy
corresponds to $\protect\omega_c$. Non-Hermitian AB caging with all bands
flat at energy $\pm 1$, $\pm \protect\sqrt{3}/2$, and $0$ occurs at the EP $\protect\mu=0$ when $\protect\gamma=1$. Other parameters are $J=\protect\kappa=1$, and the lattice size is $40$. (e) Schematic of the quasi-1D non-Hermitian cross-stitch
lattice. Each plaquette has a $\protect\pi$ synthetic
magnetic flux. The pair of CLSs is localized in a single unit cell at $\protect\gamma=\protect\kappa$. Asymmetric couplings $\protect\kappa/2+\Gamma $ in the bracket are downward for $\Gamma\neq0$. }
\label{fig1}
\end{figure}

\textit{Unidirectional coupling induced AB cage.---}In Fig.~\ref{fig1}(a),
all the primary resonators (round shape) have resonant frequency $\omega _{%
\mathrm{c}}$, evanescently coupled to the auxiliary resonators (stadium
shape). The auxiliary resonators with black arrows induce reciprocal
(symmetric) couplings between the primary resonators \cite{Hafezi} and the
auxiliary resonators with colored arrows induce asymmetric couplings \cite%
{LonghiPRBSR}. In the coupled ring resonator array, extra gain and loss
present in the half-perimeter of the auxiliary resonators. Photons tunneling
between the primary resonators through the auxiliary resonators are
amplified or attenuated and the effective coupling strength induced between
the primary resonators is tunneling direction dependent \cite{LonghiPRBSR};
consequently, extra half-perimeter gain (loss) in the auxiliary resonator
leads to the asymmetric coupling and creates an imaginary gauge field \cite%
{Hatano,Midya,XZZhang,LJin19}.

The equations of motion for the light field of the resonator array in the
coupled mode theory is equivalent to the discrete Schr\"{o}dinger equations
for the 1D lattice shown in the lower panel of Fig.~\ref{fig1}(a) \cite%
{Mukherjee15,RamezaniFB,LJinPRL,CMT}; the Hamiltonian of the schematic
lattice models the resonator array. In coupled mode theory, the equations of
motion for the light field in the chiral-inversion symmetric non-Hermitian
photonic crystal lattice for the nonzero $\Gamma $ are in the following form
\begin{eqnarray}
i\frac{\mathrm{d}\varphi _{4j}}{\mathrm{d}t} &=&-i\Gamma \varphi _{4j}+\nu
\varphi _{4j-1}+J\varphi _{4j+1},  \label{1} \\
i\frac{\mathrm{d}\varphi _{4j+1}}{\mathrm{d}t} &=&i\Gamma \varphi
_{4j+1}+J\varphi _{4j}+\nu \varphi _{4j+2},  \label{2} \\
i\frac{\mathrm{d}\varphi _{4j+2}}{\mathrm{d}t} &=&-i\Gamma \varphi
_{4j+2}+\mu \varphi _{4j+1}+J\varphi _{4j+3},  \label{3} \\
i\frac{\mathrm{d}\varphi _{4j+3}}{\mathrm{d}t} &=&i\Gamma \varphi
_{4j+3}+J\varphi _{4j+2}+\mu \varphi _{4j+4},  \label{4}
\end{eqnarray}%
which are equivalent to the discrete Schr\"{o}dinger equations of a 1D
non-Hermitian lattice Hamiltonian $H$ and $\varphi _{l}$ is the light field
amplitude (wave function) in the $l$-th primary resonator. The subscript $l$
is the index of the lattice from the left to the right of Fig.~\ref{fig1}%
(a). The equations of motion also describe the dynamics of the coupled
waveguide lattice.

The non-Hermitian Aharonov-Bohm (AB) cage forms at the exceptional point
(EP) of the 1D non-Hermitian lattice, which is also $\mathcal{PT}$-symmetric
and the EP is the $\mathcal{PT}$-symmetric phase transition point. The
Hamiltonian of the 1D non-Hermitian lattice at the EP is unable to be
diagonalized. However, the time evolution dynamics of the non-Hermitian AB
cage can be straightforwardly obtained from solving the discrete Schr\"{o}%
dinger equations, i.e., straightforwardly solving the set of differential
equations (\ref{1})-(\ref{4}). The analytical time-evolution dynamics of a
single-site initial excitation is exemplified in the eight-site 1D
non-Hermitian lattice shown in Fig.~\ref{fig1}(a).

Each dimer has a reciprocal coupling $J$. The asymmetric couplings with
opposite orientations connect every two nearest neighbor dimers; the
asymmetric inter dimer coupling strengths are $\mu =\left( \kappa -\gamma
\right) /2$ and $\nu =\left( \kappa +\gamma \right) /2$. The primary
resonators have extra gain and loss $\pm i\Gamma $. At the EP $\gamma
=\kappa $, the asymmetric couplings are unidirectional. Any excitation in
the red dimer is confined there without escaping because the outward
couplings ($\mu =0$) in magenta vanish and the red dimers only possess
inward intercell unidirectional couplings. In contrast, any excitation in
the orange cell tunnels to the two neighbor red dimers due to the
nonvanishing outward couplings ($\nu =\kappa $) in green, and then the
excitation is confined in the two neighbor red dimers.

Applying the Fourier transformation, we obtain the energy bands of the
lattice $E\left( k\right) =\pm \sqrt{J^{2}-\Gamma ^{2}+\mu \nu \pm 2J\sqrt{%
\mu \nu }\cos \left( k/2\right) }$, where the momentum is $k=2\pi n/N$ and $%
n\in \lbrack 1,N]$. The spectra of the photonic resonator lattice for $%
\Gamma =0$ and $1/2$ are presented in Figs.~\ref{fig1}(b) and \ref{fig1}(c).
The black hollow circles mark two flat bands; all the levels are two-state
coalesced at $\pm \sqrt{J^{2}-\Gamma ^{2}}$, forming the non-Hermitian AB
cage with two coalesced flat bands and resulting in the complete confinement
of light excitations. The non-Hermitian AB caging at a high-order EP of
four-state coalescence \cite{LJinPRA97} appears when the $\mathcal{PT}$%
-symmetric non-Hermitian dimers are at their own EP ($J=\Gamma $) in
addition to the unidirectional inter cell coupling ($\gamma =\kappa $);
where all the bands coalesce to a zero-energy flat band as depicted in Fig.~%
\ref{fig1}(d).

\begin{figure}[tb]
\includegraphics[ bb=0 0 550 350, width=8.8 cm, clip]{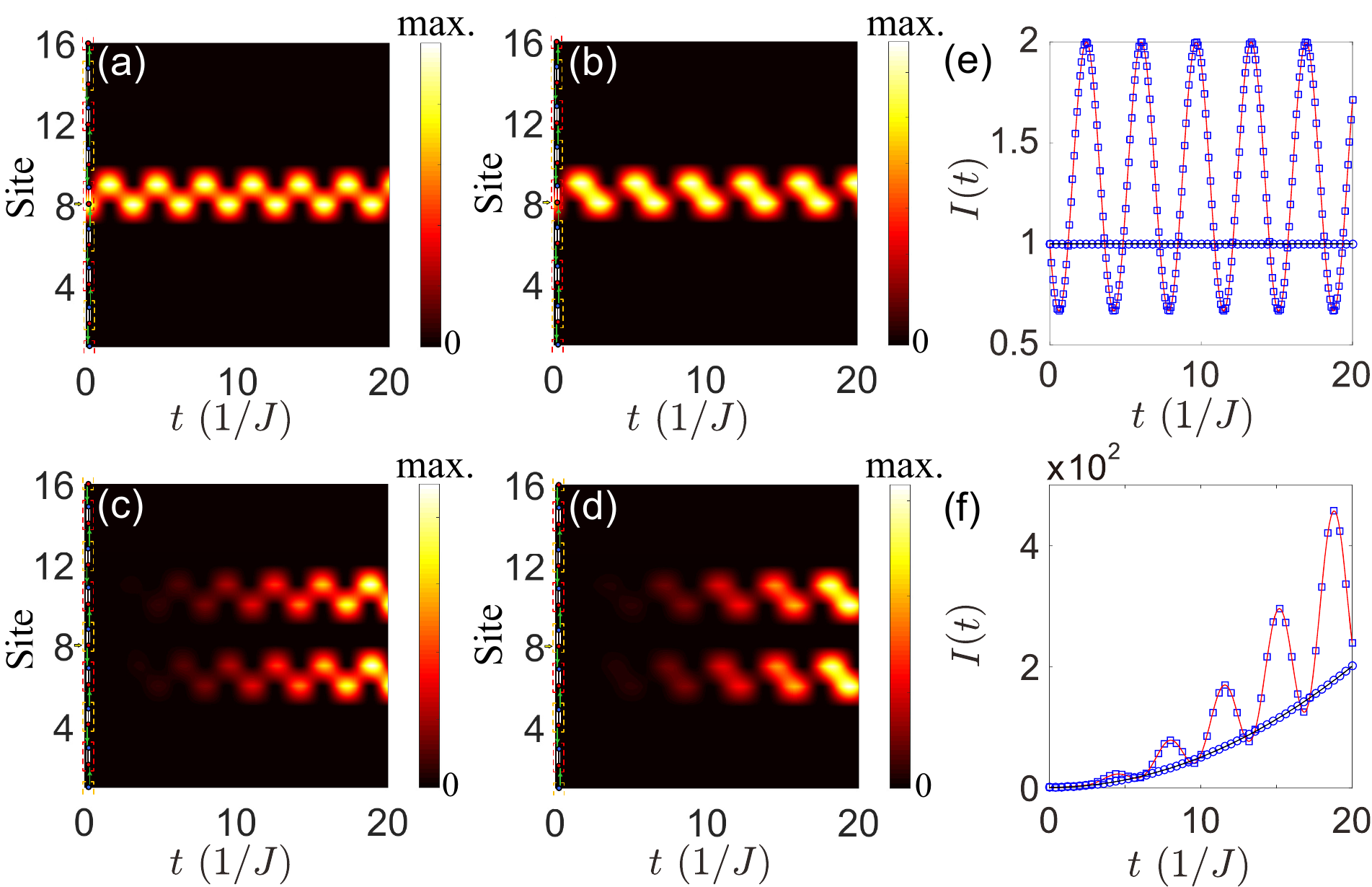}
\caption{Non-Hermitian AB caging. Initial excitation is indicated by the
yellow arrow in the schematic. The excitation is in (a, b) the red and (c,
d) the orange cells. (e) The constant (blue circle) and oscillating (blue
square) intensities are depicted for the excitation in the red cell for (a)
and (b). (f) The polynomially increasing intensities for the excitations in
the orange cell of (c) and (d) are depicted by the blue circle and blue
square, respectively. $\Gamma=0$ in (a, c) and $\Gamma=1/2$ in (b, d); other
parameters are $J=\protect\kappa=\protect\gamma=1$. The analytical
intensities are indicated by the red and black lines.} \label{fig2}
\end{figure}

The lattice Hamiltonian at the EP is defective, each two-state coalesced
eigenstate is associated with a generalized eigenstate \cite{Golub}. The
intensity of initial excitation including eigenstates with different
eigenvalues experiences an oscillation \cite{CE,JLPRA84}; and that including
the generalized eigenstate quadratically increases with time \cite{PWang,CLi}%
. The localized excitation in the non-Hermitian flat band can increase
polynomially as a result of the unique dynamics at the EP \cite%
{LeykamFB,LGePR}. Figure~\ref{fig2} shows the intensities of different
excitations.

The excitation in a red dimer has limited intensity; whereas the excitation
in an orange dimer has quadratically increasing intensity and is confined in
its two red neighbor dimers. The excitation on site $8$ ($9$) cannot tunnel
leftward (rightward) to site $7$ ($10$). The excitation on site $8$ in Fig.~%
\ref{fig2}(a) [Fig.~\ref{fig2}(b)] is relevant to the orthogonal
(nonorthogonal) eigenstates of the Hermitian (non-Hermitian) lattice; the
confined excitation inside the red dimer of sites $8$ and $9$ has constant
(oscillating) intensity \cite{LJin2011}, which is represented by the blue
circles (squares) in Fig.~\ref{fig2}(e). The non-Hermiticity results in the
nonunitary time evolution; the superposition of eigenvectors with different
energies results in the oscillation. By contrast, any excitation in the
orange dimer can only tunnel to its two neighbor red dimers, the excitations
cannot tunnel backward and then are confined there [Figs.~\ref{fig2}(c) and~%
\ref{fig2}(d)]. The excitation in Fig.~\ref{fig2}(c) [Fig.~\ref{fig2}(d)] is
relevant to the generalized eigenstates, which lead to a linear
time-dependent excitation amplitude \cite{PWang}; the excitation intensity
increases quadratically over time as presented in Fig. \ref{fig2}(f) by the
blue circles (squares). For the non-Hermitian AB caging at the high-order
EP, the excitation intensity increases more rapidly~\cite{Ganainy18}.

\textit{Destructive interference induced AB cage.}---Synthetic magnetic
field for photons has been realized and boosts the study of topological
photonics \cite%
{BlochReview,Goldman,SLZhu,Cooper,Goldman14,Xiao,ZY,LLu,TOzawa,Rechtsman},
particularly in photonic crystals of coupled resonators \cite%
{Hafezi,HafeziNP2011,QLin} and waveguides \cite{Yu,Fang}. The Peierls phase
induced by the dynamic modulation \cite{Yu,Fang,Roushan,GoldmanPRX} or path
length imbalance method in photonics \cite{Hafezi,HafeziNP2011} generates
the synthetic magnetic flux, which helps the realization of an AB cage in
the Hermitian rhombic lattice \cite%
{SLOL,Goldman18,ASarXiv,Maluckov,DiLiberto}.

The non-Hermitian AB cage under the synthetic magnetic flux is based on
another confinement mechanism of the destructive interference. In Fig.~\ref%
{fig1}(e), the nonreciprocal coupling $\pm iJ/2$ has a Peierls phase factor $%
e^{\pm i\pi /2}$, which leads to a $\pi $ magnetic flux in each plaquette of
the cross-stitch lattice. The destructive interference results in the CLSs
localized in a single plaquette and the wavefunctions for the coalesced flat
band energies $\pm J$ are marked in the lower panel. In the discrete Schr%
\"{o}dinger equations, the nonreciprocal coupling and the cross-stitch
coupling multiplying the wavefunctions $\left( 1,i\right) ^{\mathrm{T}}$
cancel at the top-left site $\left( -iJ/2\right) \cdot \left( 1\right)
+\left( J/2\right) \cdot \left( i\right) =0$ and at the bottom-left site $%
\left( J/2\right) \cdot \left( 1\right) +\left( iJ/2\right) \cdot \left(
i\right) =0 $; similarly, the interferences of the wavefunctions at the
top-right and bottom-right sites are fully destructive. The destructive
interference results in zeros for the wavefunction outside the plaquette
indicated by the black dashed square.

The synthetic magnetic field associated with balanced gain and loss relates
to the asymmetric coupling. Applying a unitary transformation to the $4N$%
-size 1D non-Hermitian lattice in Fig.~\ref{fig1}(a), we obtain the quasi-1D
non-Hermitian cross-stitch lattice in Fig.~\ref{fig1}(e). The dynamics of
the non-Hermitian cross-stitch lattice is straightforwardly obtained from
the dynamics of the 1D non-Hermitian lattice through the unitary
transformation.

\begin{figure}[tb]
\includegraphics[ bb=0 0 590 500, width=8.8 cm, clip]{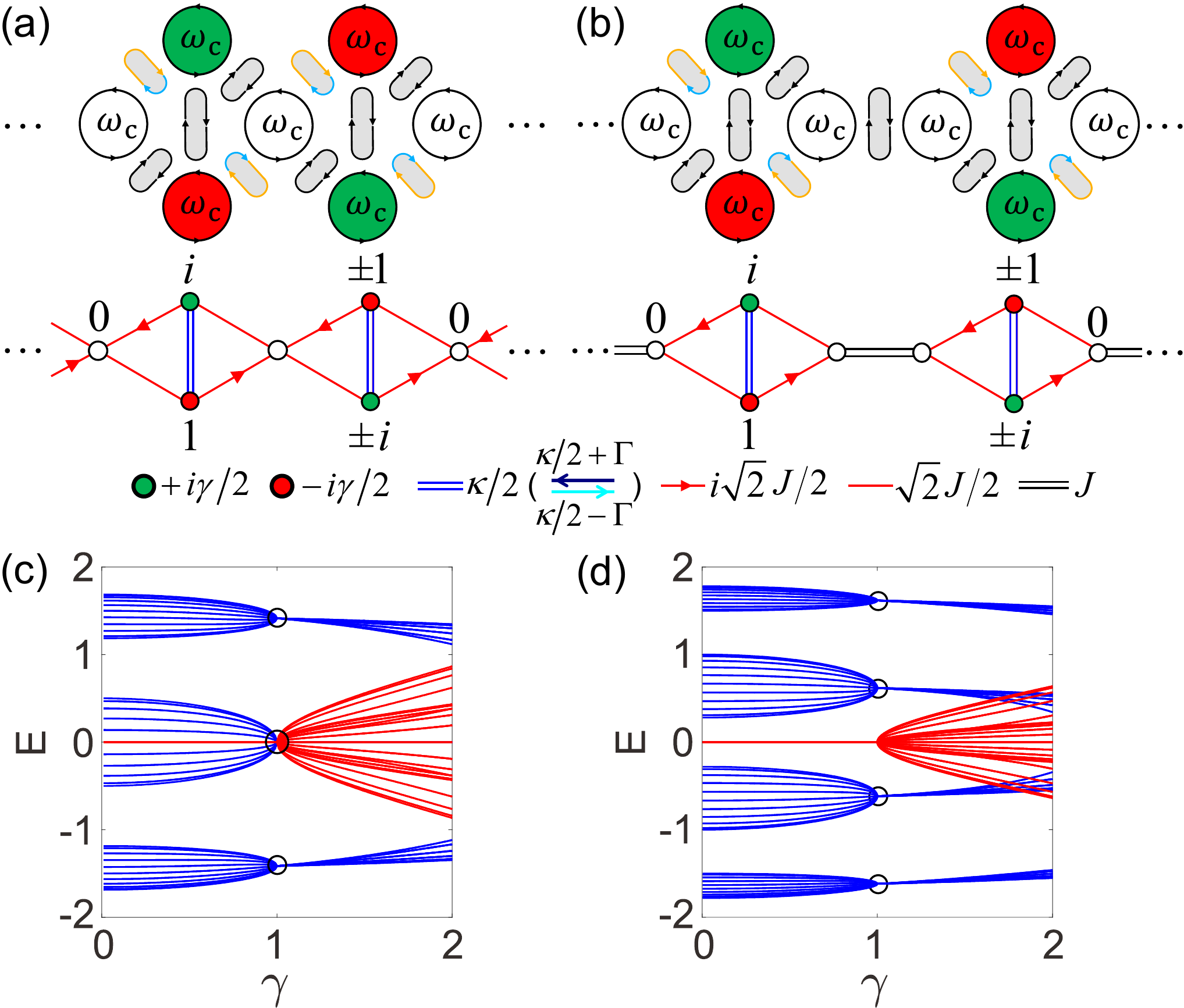}
\caption{(a, b) Schematic of the quasi-1D non-Hermitian rhombic lattices
under inversion symmetry ($\Gamma=0$). Each triangular area has a $\protect\pi/2$ synthetic magnetic flux as a consequence of the Peierls phase induced
by the optical length imbalance indicated in the orange and cyan arrows~\protect\cite{Hafezi}. (c) [(d)] Spectrum of $60$ ($80$) sites for (a)
[(b)], the real (imaginary) part is in blue (red). The parameters are $\Gamma=0$, $J=\protect\kappa=1$. The non-Hermitian AB cages of three and
four flat bands appear at the EP $\protect\gamma=1$.} \label{fig3}
\end{figure}

\begin{figure*}[t]
\includegraphics[ bb=0 0 590 200, width=18.0 cm, clip]{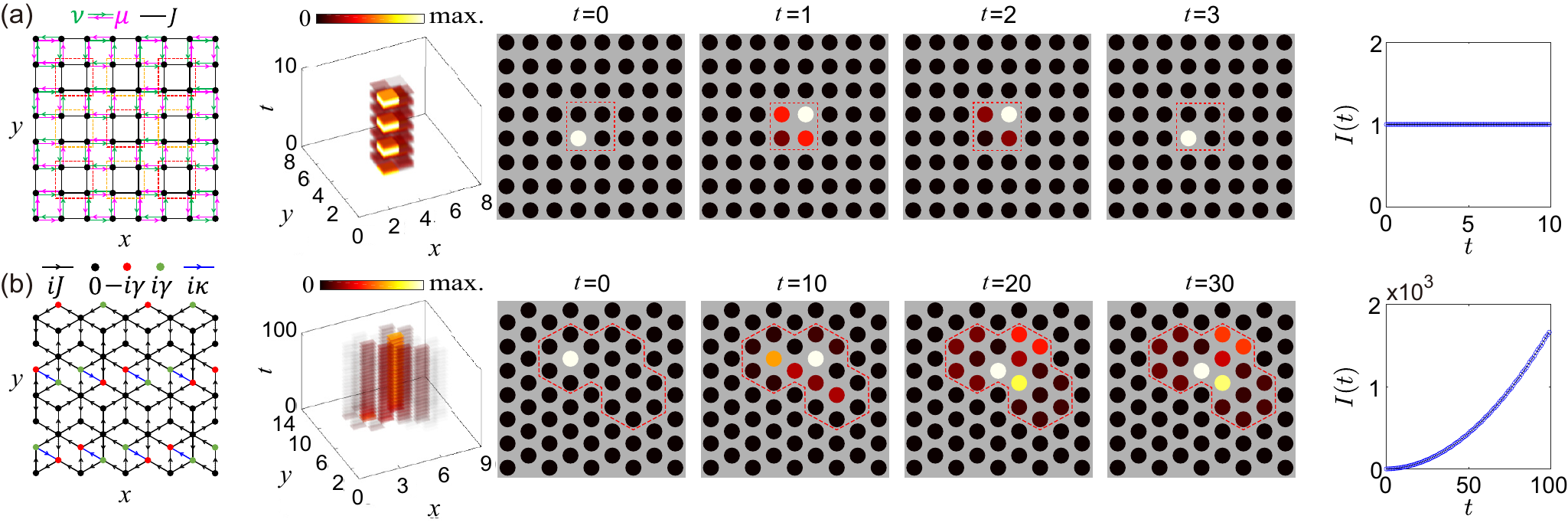}
\caption{Non-Hermitian AB caging in the 2D non-Hermitian (a) square lattice
and (b) dice lattice. The panels from the left to the right are the lattice
schematics, the excitation intensity distribution, and the total excitation
intensity for a single-site excitation. Light field is confined inside the
dashed red polygon. The parameters are (a) $J=1$, $\protect\mu=0$, $\protect%
\nu=1$; (b) $J=1$, $\protect\kappa=\protect\gamma=2$.}
\label{fig4}
\end{figure*}

The oppositely oriented unidirectional couplings can form non-Hermitian AB
cages consisting of multiple flat bands if we consider oligomer cells
instead of dimer cells in Fig.~\ref{fig1}(a). The quasi-1D non-Hermitian
rhombic lattices with synthetic magnetic flux and balanced gain and loss
illustrated in Figs.~\ref{fig3}(a) and~\ref{fig3}(b) can be obtained
similarly under the unitary transformation applied to the connecting sites
associated with the unidirectional couplings; the corresponding energy bands
are shown in Figs.~\ref{fig3}(c) and~\ref{fig3}(d), where the non-Hermitian
AB caging appears at the EP $\gamma =\kappa $.

The Hermitian rhombic lattice in Fig.~\ref{fig3}(a) at $\gamma =\kappa =0$
is experimentally realized in the coupled optical waveguides \cite%
{Goldman18,ASarXiv}. Alternatively introducing $\mathcal{PT}$-symmetric
dimers at the EP in the rhombic lattices forms the non-Hermitian AB cages
[Figs.~\ref{fig3}(a) and~\ref{fig3}(b)]. The eigenstate of the $\mathcal{PT}$%
-symmetric dimer at the EP $\gamma =\kappa $ is $\left( i,1\right) ^{\mathrm{%
T}}$, the contributions of the extra non-Hermiticity $\gamma $ and the
coupling $\kappa $ in the discrete Schr\"{o}dinger equations cancel for the
site with gain $\left( \kappa /2\right) \cdot \left( 1\right) +\left(
i\gamma /2\right) \cdot \left( i\right) =0$ and for the site with loss $%
\left( \kappa /2\right) \cdot \left( i\right) +\left( -i\gamma /2\right)
\cdot \left( 1\right) =0$. Thus, the flat band energies and the eigenstates
are unchanged in the presence of $\mathcal{PT}$-symmetric dimers at the EP,
but become coalesced energy levels and coalesced CLSs.

The lower panel of Fig.~\ref{fig3}(a) illustrates the cage solutions of the
non-Hermitian rhombic lattice. The flat band energy is $0$ ($\pm \sqrt{2}J$%
), the corresponding CLS amplitude of the unmarked hollow circle in the
center is $0$ ($\pm 2i$). The wavefunctions of the left (right) dimer
destructively interferes at the left (right) site, resulting in the marked
zeros. When the symmetric coupling $\kappa /2$ becomes asymmetric coupling $%
\kappa /2\pm \Gamma $, all six bands coalesced to a single zero-energy flat
band at $\Gamma =\sqrt{2}J$.

\textit{Non-Hermitian AB cages in 2D.}---The generality of the two
confinement mechanisms is further elucidated through the proposed 2D
non-Hermitian AB cages. Figure~\ref{fig4}(a) is the schematic of the
non-Hermitian square lattice that exploits the unidirectional coupling to
realize the light confinement. Figure~\ref{fig4}(b) is the schematic of the
non-Hermitian dice lattice that supports the light confinement through the
destructive interference under the appropriate match between synthetic
magnetic flux and non-Hermiticity.

At the EP $\mu =0$ of the non-Hermitian square lattice, all the bands are
flat and the energies are $0,\pm 2J$. The excitations are confined inside
the red plaquette with four symmetric couplings $J$ in black. The flow of
light enters the red plaquette unidirectionally as a consequence that the
inter-plaquette couplings of the red plaquettes are inward. Photons tunnel
to the red plaquettes without escaping. For any excitation in the orange
plaquette, photons tunnel to their neighbor red plaquettes and are confined
there.

In the absence of the non-Hermiticity $\pm i\gamma $ and the nonreciprocal
coupling $i\kappa $, Fig.~\ref{fig4}(b) reduces to the Hermitian dice
lattice, which is also referred to as the $\mathcal{T}_{3}$ lattice \cite%
{Vidal}. The synthetic magnetic flux in every diamond plaquette of the dice
lattice is $\pi $ and the destructive interference results in the AB caging
with all bands flat at energies $0,\pm \sqrt{6}J$. The synthetic magnetic
flux $\pi $ in each diamond plaquette is separated by $i\kappa $ into two $%
+\pi /2$ or $-\pi /2$ in the triangular areas. The synthetic magnetic fluxes
associated with the gain and loss maintains the destructive interference;
the eigenvalues and the eigenstates of the Hermitian dice lattice are
unchanged after introducing the $\mathcal{PT}$-symmetric dimers at the EP $%
\gamma =\kappa $. The contributions of the $\mathcal{PT}$-symmetric dimers
cancel in the discrete Schr\"{o}dinger equations of the dice lattice.
Arbitrary local light excitation is entirely confined. Notably, the
non-Hermitian generalization is nontrivial, we emphasize that although the
CLSs are \textit{unchanged} by the non-Hermiticity; the localization of
excitation \textit{changes} due to the CLSs coalescence and the EP
unidirectionality \cite{CLi}. For comparison, the excitation localizes in
single snowflake in the Hermitian dice lattice.

The dynamics of a single-site initial excitation in the 2D non-Hermitian AB
cages is simulated and presented in the middle panels of Fig.~\ref{fig4}.
The initial excitation is localized in the plaquette of the non-Hermitian
square lattice [Fig.~\ref{fig4}(a)] and localized in the three hexagonal
area because the snowflake shape CLSs for the non-Hermitian dice lattice
[Fig.~\ref{fig4}(b)]. The single-site excitation in Fig.~\ref{fig4}(b)
relates to the generalized eigenstate of the system at the EP; consequently,
the excitation probability polynomially increases with time.

\textit{Conclusion.}---We have proposed non-Hermitian AB cages in 1D and 2D
at the EP and demonstrated two confinement mechanisms. (i) The
unidirectional couplings imposed the inversion (combined-inversion)
symmetry. (ii) The destructive interference at the interplay between the
gain and loss, the Peierls phase, and the couplings. In both mechanisms, the
additional non-Hermitian elements at the EP retain the eigenstates
localization; any light excitation not limited to the eigenstate of the
non-Hermitian AB cage is completely confined. However, the spectrum of the
non-Hermitian AB cage becomes fully constituted by the coalesced flat bands;
and the localized area of excitation may alter and affected by the EP
unidirectionality. The non-Hermiticity induces the nonunitary dynamics. The
intensity may periodically oscillate or polynomially increase. The
non-Hermitian AB cages proposed in the photonic resonator lattices can be
implemented in the coupled waveguides as well~\cite{SzameitPRL19,Goldman18}.

In addition, offsetting an overall imaginary energy $i\gamma $ in the
passive system with losses $0$ and $-2i\gamma $ yields the active $\mathcal{%
PT}$-symmetric lattice with balanced gain and loss $\pm i\gamma $~\cite%
{Poli,Weimann,MPan,Rechtsman19,SzameitPRL19,SzameitNC19}; the non-Hermitian
AB cages are possible to be realized in passive photonic lattice through
judiciously engineering losses of the sublattices. Recent progresses on the
Hermitian AB cage \cite{Goldman18}, the non-Hermitian flat band \cite%
{SzameitPRL19}, and the $\mathcal{PT}$-symmetric photonic crystal \cite%
{SzameitNC19,Rechtsman19} indicate the non-Hermitian AB cages proposed are
experimentally accessible.

\textit{Acknowledgement.---}We acknowledge the support from National Natural
Science Foundation of China under Grant No.~11975128.

\end{document}